\documentstyle[11pt,newpasp,twoside,epsf]{article}
\def\degree{$^{\circ}$}
\def\kms{km~s$^{-1}$}
\def\ga{\mathrel{\hbox{\rlap{\hbox{\lower4pt\hbox{$\sim$}}}\hbox{$>$}}}}
\def\la{\mathrel{\hbox{\rlap{\hbox{\lower4pt\hbox{$\sim$}}}\hbox{$<$}}}}

\begin{document}

\title{Short-Spacings Correction from the Single-Dish Perspective}
\author{Snezana Stanimirovic}
\affil{National Astronomy and Ionosphere Center, Arecibo Observatory, HC 3
Box 53995, Arecibo, Puerto Rico 00612, USA}

\begin{abstract}
While, in general, interferometers provide high spatial resolution for
imaging small-scale structure (corresponding to high spatial frequencies in
the Fourier plane), single-dishes can be used to image the largest spatial
scales (corresponding to the lowest spatial frequencies), including the
total power (corresponding to zero spatial frequency). For many
astrophysical studies, it is essential to bring `both worlds' together by
combining information over a wide range of spatial frequencies. This
article demonstrates the effects of missing short-spacings, and
discusses two main issues: (a) how to provide missing short-spacings
to interferometric data, and (b) how to combine short-spacing single-dish
data with those from an interferometer.
\end{abstract}

\section{Introduction}
\label{s:intro}

All radio telescopes can be classified as either filled or
unfilled-aperture antennas. The simplest 
filled-aperture antennas are single-dish telescopes.
The desired astronomical object to be observed and
the particular scientific goals determine which type of radio telescope
to use.  
In general, single-dishes are considered as tools for low spatial
resolution observations, while interferometers are used for high resolution
observations. While compact objects are more suited for interferometric
observations, extended objects are commonly observed with
single-dishes as interferometers cannot faithfully recover information 
on the largest spatial scales.  

However, in many scientific cases it is essential to obtain high spatial 
resolution observations  of large objects, and to accurately represent emission 
present over a wide range of spatial scales. 
A simple recipe you may follow in such cases is:
\begin{itemize}
\item observe (mosaic) your object with an interferometer,
\item observe your object with a single-dish, 
\item cross-calibrate the two data sets, and then
\item combine the single-dish and interferometer data. 
\end{itemize}
This combination of single-dish and interferometer data, when 
observing extended objects, is referred to as the short-spacings correction.

This `simple' recipe may be considered as an artistic touch to the
interferometric images as it makes them look much nicer but still preserves
their high spatial resolution. This results from: (a) inclusion of more 
resolution elements, those seen by a single-dish; and (b) reconstruction of
image artifacts. At the same time, these images contain information about
the total power,  and can be used to measure accurate flux densities,
column densities, masses, etc. From a pure historical perspective, 
the short-spacings correction  bridges the gap 
between the two classes of radio telescopes, 
essentially obtaining the best of `both worlds', that is the high spatial
resolution information provided by interferometers, and the low spatial
resolution, including the total power, information provided by
single-dishes. 

This article will explain what the short-spacing problem is, how
it is manifested, and how we can, both theoretically and practically, solve
this problem. Section~2 depicts very briefly the 
fundamentals of interferometry, defines the spatial frequency domain, and
draws an analogy between a single dish and an
interferometer. The effects of missing short spacings are
demonstrated in Section 3, as well as prospects for solving
the problem. Section~4 considers the cross-calibration of
interferometer and single dish data which is a precursor to any combination
method. Methods for data combination are discussed in
Section~5 and Section 6, and compared 
in Section~7.

\section{A Very Brief Introduction to Interferometry}
\label{s:sf-domain}

\begin{figure}
\caption{\label{f:interferometer} {\bf [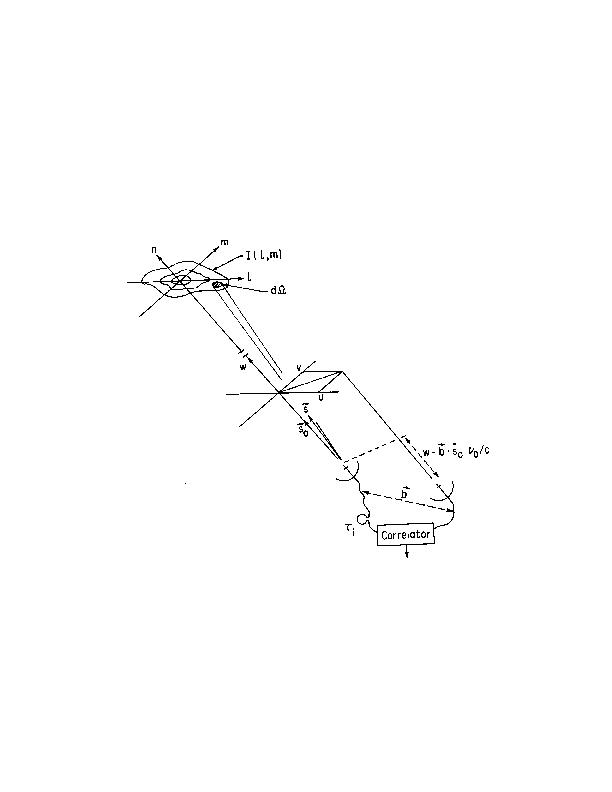]}A schematic 
diagram of the spatial
frequency and image coordinate systems. The spatial frequency domain,
$(u,v,w)$, is used to express the interferometer baseline, and 
the image domain, $(l,m,n)$, is used to express the source 
brightness  distribution.  The direction to the center of
the field of view is given by ${\bf s}_{0}$, and to any given position
by ${\bf s}$. Hence, ${\bf s} = {\bf s_{0}} +  \mbox{\boldmath{$\sigma$}}$. (From
Thompson 1994.) }
\end{figure}

A very brief review of the basics of interferometry is necessary right
at the beginning of this article, in order
to define and explain some terms that will be used further on.
However, we do not want to go deeply into interferometry, as there is a 
vast literature available on this topic, starting with 
``Interferometry and Synthesis in Radio Astronomy''
(Thompson, Moran, \& Swenson 1986) and ``Synthesis Imaging in Radio
Astronomy'' (Taylor, Carilli, \& Perley 1999).
 
The fundamental idea behind interferometry is that a Fourier transform
relation exists between the sky radio brightness distribution $I$ and the
response of a radio interferometer. 
If the distance between two antennas (the baseline) is ${\bf d}$,
then the so-called  visibility function, $V({\bf d})$, is given by:

\begin{equation}
\label{e:int-basic}
V({\bf d})= \int_{\rm source} A(\mbox{\boldmath{$\sigma$}}) I( 
\mbox{\boldmath{$\sigma$}}) \exp \left [-2\pi i ~{\bf d} \cdot 
\mbox{\boldmath{$\sigma$}} /\lambda \right] d\Omega \;.
\end{equation}
Here, $A(\mbox{\boldmath{$\sigma$}})$ is an antenna reception pattern, or 
{\bf primary beam}, and $\mbox{\boldmath{$\sigma$}}$ is
the vector difference between a given celestial position and the central 
position of the 
field of view. The {\bf aperture synthesis technique} is a 
method of solving Equation~\ref{e:int-basic} 
for $I(\mbox{\boldmath{$\sigma$}})$ by 
measuring $V$ at suitable values of ${\bf d}$. 

To simplify Equation~\ref{e:int-basic}, a more convenient, right-hand 
rectilinear, coordinate 
system is introduced in Fugure~\ref{f:interferometer}. Coordinates 
of vector ${\bf d}$ in this system are $(u,v,w)$, 
where the direction to the source center ${\bf s}_{0}$ defines
the $w$ direction, and $u$ and $v$ are baseline projections onto 
the plane perpendicular to the ${\bf s}_{0}$ direction, towards the East and
the North, respectively.
A synthesized image in the $l-m$ plane represents a projection of the
celestial sphere onto a tangential plane at the source center. 
In certain conditions, that is in the case of an Earth tracking, 
East-West interferometer array, with the $w$-axis lying in the 
direction of the celestial pole, further simplifications of
Equation~\ref{e:int-basic} are possible:
\begin{equation}
\label{e:v}
V(u,v)= \int \int A(l,m)I(l,m) \exp [-2\pi i (ul+vm)] \frac{dl dm}
{\sqrt{1-l^{2}-m^{2}}} \;.
\end{equation}
Therefore, the visibility function $V(u,v)$ can be expressed as the Fourier 
transform of a modified brightness distribution $A(l,m)I(l,m)$.  
Coordinates $u$ and $v$ ($w=0$) are measured in units of wavelength and the
$u-v$ plane is called {\bf the spatial frequency domain}.
These are effectively projections of a
terrestrial baseline onto a plane perpendicular to the source direction.
The $l-m$ plane is referred to as {\bf the image domain}.
To obtain $I(l,m)$, from Equation~\ref{e:v}, an inverse Fourier 
transform of $V(u,v)$ is required,
meaning that a complete sampling of the spatial frequency domain
is essential. In practice however, a bit more than a simple inversion is
needed as only limited sampling of the $u-v$ plane is available.

\subsection{How do we `fill in' the $u-v$ plane?}

\begin{figure}
\caption{\label{f:tracks12} {\bf [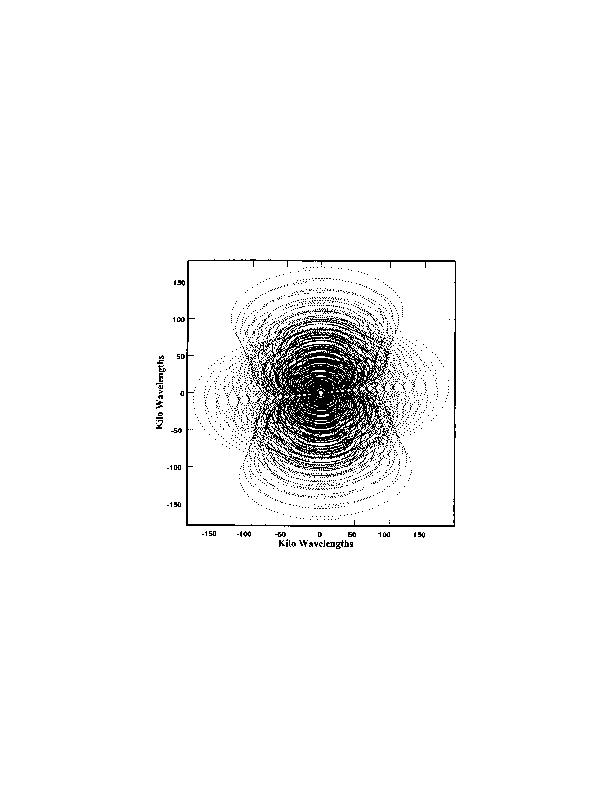]}The $u-v$ coverage 
for an 8 hr tracking
observation with the VLA, for an object at declination of 30\degree~(from
Burke \& Graham-Smith 1997).}
\end{figure}

For a given configuration of antennas any interferometer array has a limited
range of baselines, lying between a minimum,  $d_{\rm min}$\footnote{The
shortest baseline that can be achieved is constrained by the physical
limitations in placing two dishes together and the shadowing effect of one
dish by another one.}, and maximum, $d_{\rm max}$, baselines.
As an example, 5 antennas of
the Australia Telescope Compact Array (ATCA)  form 10 baselines, with
$d_{\rm min}=31$ m and $d_{\rm max}=459$ m for a particularly compact
configuration. The final resolution ($\theta_{\rm int}$) is inversely
related to the maximum baseline by  $\theta_{\rm int} \approx \lambda/ d_{\rm
max}$.  In the case of an Earth tracking interferometer, as the Earth
rotates the baseline projections on the $u-v$ plane trace a series of
ellipses.   The parameters of each ellipse depend upon the declination of
the source, the length and orientation of the baseline, and the latitude of
the center of the baseline (Thompson et al.  1986).  The ellipses are
concentric for a linear array  (e.g. ATCA). For  a 2-dimensional 
array (e.g. the Very Large Array, VLA), the ellipses
are not concentric and so can intersect.

As each baseline traces a different ellipse, the ensemble of ellipses
indicates the spatial frequencies that can be measured by the array (see
Thompson et al. 1986).  At each sampling interval, the correlator measures
the visibility function for each baseline,  thus resulting in a
number of samples being measured over elliptical tracks in the $u-v$ plane.
Hence, the resultant  interferometer  $u-v$ coverage will always be, more
or less, incomplete, having a hole in the center of the $u-v$ plane whose
diameter corresponds to the minimum baseline, gaps between measured
elliptical tracks, and gaps between each adjacent samples on each
elliptical track.  The ensemble of ellipses (loci) is known as the {\bf
transfer} or {\bf sampling function}, $b_{\rm int}(u,v)$. An example of the
sampling function obtained with the VLA is shown in Fugure~\ref{f:tracks12}.

Hence, if $V(u,v)$ is a true (ideal) visibility function, 
the measured (observed) visibilities ($V_{\rm int}'$) can be expressed as:
\begin{equation}
\label{e:int1}
V'_{\rm int}(u,v)=V(u,v)b_{\rm int}(u,v) \;.
\end{equation} 
$b_{\rm int}$ is usually representable
by a set of $\delta$-functions, between the lowest and the highest
spatial frequency sampled by the interferometer (corresponding to the
shortest and the longest baselines, respectively).  The Fourier transform
of Equation~\ref{e:int1} gives the observed sky brightness 
distribution $I_{\rm int}^{\rm D}$ (so called {\bf `dirty' image}):
\begin{equation}
\label{e:int}
I_{\rm int}^{\rm D}(l,m) = I(l,m)* B_{\rm int}(l',m')\;,
\end{equation}
where $B_{\rm int}$ is {\bf the synthesized or `dirty' beam}, which is 
the point source response of the interferometer. As usually, 
asterisks ($*$) are used to denote convolution.
When imaging, incomplete $u-v$ coverage leads to severe artifacts, such
as negative `bowls' around emission regions and negative and positive
sidelobes (Cornwell, Braun, \& Briggs 1999). We return to this in
Section~\ref{s:effects} The determination of $I$ from $I_{\rm int}^{\rm D}$
in the deconvolution process, requires beforehand interpolation and
extrapolation of $V_{\rm int}'$ for missing data due to the discontinuous
nature of $b_{\rm int}$ (Cornwell \& Braun 1989).
This process works well when a compact configuration of antennas is used
and when the source is small enough, with angular size 
$\theta \leq 2\lambda/d_{\rm min}$ (Bajaja \& van Albada 1979). 

For imaging larger objects, with angular size
$\theta > \lambda/d_{\rm min}$, a significant improvement  in filling in
the $u-v$ coverage can be  achieved by using the {\bf `mosaicing
technique'}, where observations of many pointing centers are obtained and
`pasted' together (see Holdaway 1999).  
Mosaicing effectively reduces the shortest projected baseline to 
$d_{\rm min}-D/2$, where $D$ is the diameter of an individual antenna. 
Nevertheless, the center of the $u-v$
plane still suffers if significant large scale structure is present. This
lack of information for very low  spatial frequencies (around the center of
the $u-v$ coverage) in  an interferometric observation is usually referred
to as  the {\bf `short-spacings problem'}.
 
\subsection{A single-dish as an interferometer?}

\begin{figure}
\plotone{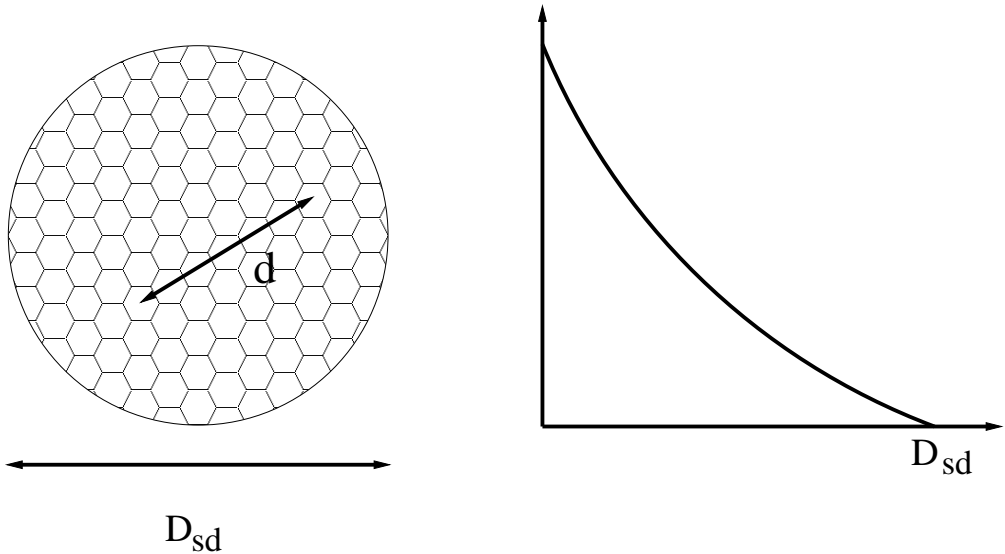}
\caption{\label{f:dish}A schematic representation of a single dish as an
interferometer with a large number of elements and a monotonically decreasing
 distribution of baselines ($d$) from zero to $D_{\rm sd}$.}
\end{figure}

Extended objects with angular size $\theta > \lambda/d_{\rm min}$  can be
observed with a single-dish.   Let us now think of a single-dish in a
slightly unusual way, imagining filled apertures consisting of a
large number of small panels  packed closely together. Then all these
panels can act as interferometer elements with their  signals being
combined together at the focus, making a so called phased or adding  array (see
contribution by D. Emerson in this volume).  The distance between each two
panels corresponds to a baseline, as shown in Fugure~\ref{f:dish}.  The
baseline distribution then monotonically decrease from zero at  the center
up to the maximum baseline, determined by the single-dish diameter $D_{\rm
sd}$. This is also shown in Fugure~\ref{f:dish}.

One observation with a single-dish provides a total flux density measurement,
corresponding to the zero spacing, $(u,v)=(0,0)$. However, if a single-dish
scans across an extended celestial object, it measures not only a single
spatial frequency, but a whole range of continuous spatial frequencies all
the way up to a maximum of  $D_{\rm sd}$ (Ekers \& Rots 1979). Hence,
a single-dish behaves as an  interferometer with an almost infinite number
of antennas, and therefore  has a continuous range of baselines, from zero
up to $D_{\rm sd}$.  The nice thing about this representation is that we
can now use the  same mathematical notation to describe both single-dishes
and  interferometers.

The observed sky brightness distribution $I_{\rm sd}^{\rm D}$ in 
the case of single-dish observations is then given by:
\begin{equation}
\label{e:sd1}
 I_{\rm sd}^{\rm D}(l,m) = I(l,m) * B_{\rm sd}(l',m'), 
\end{equation}
with $B_{\rm sd}$ being the {\bf single-dish beam} pattern. The Fourier 
transform of Equation~\ref{e:sd1} gives the observed single-dish 
`visibilities', $V_{\rm sd}'$:
\begin{equation}
\label{e:sd2}
V_{\rm sd}'(u,v) = V(u,v) \times b_{\rm sd}(u,v)
\end{equation}
where $b_{\rm sd}$ is the Fourier transform of the single-dish beam
pattern which, unlike $b_{\rm int}$, is a continuous function 
between zero and the highest
spatial frequency sampled by the single-dish. Determination of $I$ from
$I_{\rm sd}^{\rm D}$ requires deconvolution, but no interpolation 
of $V_{\rm sd}'$ is needed since this is a continuous function.

\begin{figure}
\caption{\label{f:effect2}  {\bf [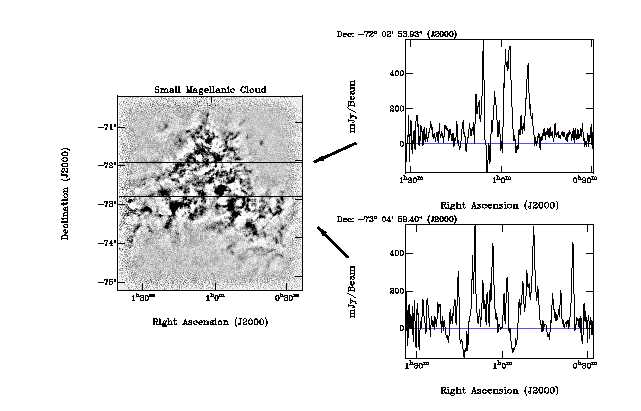]}An HI 
image of the SMC from Stanimirovic et
al. (1999) is shown on the left. The observations were obtained as a
mosaic of 320 different pointing centers with the ATCA.  Positive intensity
values are shown in black, while white  represents negative pixel
values. RA cuts through the image, at  Dec $-72^{\circ}$ $02'$ $54''$ and
Dec $-73^{\circ}$ $05'$, are shown on the right. }
\end{figure}

\section{How is the Short-spacings Problem Manifested?}
\label{s:effects}

As shown in Equation~\ref{e:int}, the sky brightness distribution can be
reconstructed, in the case of interferometric observations, by deconvolving
the `dirty' image with the synthesized beam. As an example,
Fugure~\ref{f:effect2} shows the result of the HI `mosaic' observations  of
the Small Magellanic Cloud (SMC) with the ATCA. More information  about
these observations and data reduction is available in Stanimirovic et
al. (1999). The two adjacent panels on the right side show right 
ascension (RA) cuts through the image.  Negative bowls 
(shown in white on the image) are seen around  emission peaks 
(shown in black), as well as in RA
cuts. These are  typical interferometric artifacts resulting from an
incomplete $u-v$ coverage.

\begin{figure}
\caption{\label{f:effect1} {\bf [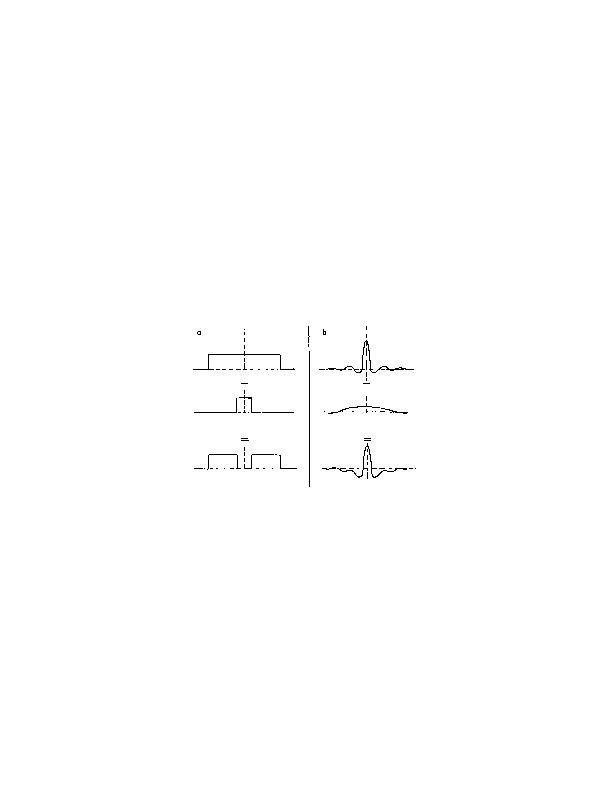]}
A 1-dimensional cut through $(u,v)=(0,0)$ of the spatial frequency domain, 
(a), and its corresponding manifestation in the image domain, (b).
An exclusion of the central values from the spatial  frequency domain
is equivalent to a subtraction of a broad pedestal in  the 
image domain, resulting in the presence of a deep negative `bowl'  
around the observed object.
(From Braun \& Walterbos 1985.)}
\end{figure}

A simple graphical explanation of why this happens, borrowed from  Braun \&
Walterbos (1985), is shown in  Fugure~\ref{f:effect1} for the case of a
point source.  The solid vertical
line in Fugure~\ref{f:effect1} distinguishes  the spatial frequency (a) 
from the image domain (b).  The distribution of measured 
spatial frequencies, or what we
have already defined as a transfer (or sampling) function, $b_{\rm int}$,
is given on the left side, while its Fourier transform, that is the
synthesized beam, $B_{\rm int}$, is shown in the right. An exclusion of the
central values from the spatial  frequency domain, is
equivalent to a subtraction of a broad pedestal in  the image domain, resulting
in the presence of a deep negative `bowl'  around the observed object, as
seen in Fugure~\ref{f:effect2}.

This demonstrates simply how severe the effects of missing  short-spacings
can be. The larger the object is relative to the  reciprocal of the
shortest measured baseline one tries to image, the more prominent the
short-spacing problem becomes.

\subsection{How can we solve the short-spacings problem?}

There are two main questions concerning the short-spacings problem:
\begin{enumerate}
\item how to provide (observe) missing short spacings
to interferometric data;
\item how to combine short-spacing
data with those from an interferometer.
\end{enumerate}
 
In answering the first question, all solutions can be grouped into two
array schemes: {\bf homogeneous}, having all antennas of the same size, and
{\bf heterogeneous}, based on observations obtained with different-sized
antennas. There are many possibilities concerning the heterogeneous arrays,
such as using smaller arrays and even a hierarchy of smaller arrays. The
simplest option, though, is a single-dish telescope with a diameter
($D_{\rm sd}$) larger than the interferometer's minimum baseline.

We briefly touch here on some pros and cons for both array schemes.
One of the difficulties in providing short-spacings with a single-dish is
that it is hard
to provide a large single-dish which would have the sensitivity equivalent
to that of an interferometer (Holdaway 1999). Also, single-dish 
observations are complex (they require a lot of separate pointing 
centers to cover a large object)
and very sensitive to systematic errors. Using theoretical analysis,
numerical simulations and observational tests,
Cornwell, Holdaway, \& Uson (1993) show that a homogeneous array in which the
short-spacings are obtained from single antennas of the array, allows high
quality imaging. They find that a key advantage over the large single-dish
scheme is pure simplicity, which is an important factor for the complex
interferometric systems. 
As both interferometric and total power data are obtained with the same
array elements, no cross-calibration is required in this case.
Note that in this case total-power and interferometric observations 
have to be synchronized which is not a simple task because of different
observing techniques involved (e.g. the single-dish observations require
frequency or position switching modes!).  This turns out to be
an especially difficult task for the continuum observations.

However, to {\it fully} fill in the central gap in an
interferometer $u-v$ coverage and {\it preserve sensitivity} at 
the same time, the heterogeneous array scheme appears more 
advantageous. This has been recently recognized in the planning of
the future Atacama Large Millimeter/Submillimeter Array (ALMA). Imaging
simulations have shown that antenna pointing errors of only a few percent
of the primary beam width produce large errors in the visibilities in the
central $u-v$ plane, causing a large degradation of image quality (see
Morita 2001). To compensate for this problem, an additional, smaller array of 
6 -- 8 m dishes, the so called ALMA Compact Array (ACA), has been 
proposed to provide short baselines.

In answering the second question, methods for the combination of
interferometer and single-dish data can be grouped into two classes: data
combination in the spatial frequency domain
(Bajaja \& van Albada 1979; Vogel et al. 1984; Roger et al. 1984;
Wilner \& Welch 1994; Zhou, Evans, \& Wang 1996), and data
combination in the image domain
(Ye \& Turtle 1991; Stewart et al. 1993; Schwarz \& Wakker 1991; Holdaway 
1999). Each approach can be realized through a number of different methods.
Both approaches are very common and are becoming a standard data processing
step. 

As the most common scheme of a heterogeneous array involves use of a
large single-dish telescope, we proceed to consider this 
particular case further.

\section{Cross-calibration of Interferometer and Single-dish Data}
\label{s:scaling}

Before adding short-spacing data, it is necessary to be sure that both
the interferometer and single-dish data sets have identical flux density
scales.  As calibration is
never perfect, the calibration differences between the two sets of observations
can be significant in some cases (e.g. observations spread over a long 
period of time, different data quality, use of different flux density scales 
for calibration, quality of calibrators, etc.). This results in 
a small but appreciable difference in the measured flux densities.

\begin{figure}
\plotfiddle{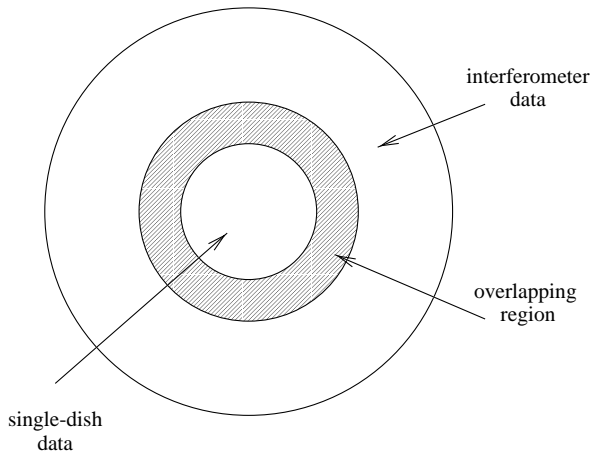}{7cm}{0}{60}{60}{-110}{0}
\caption{\label{f:uv-scheme} 
A schematic diagram of 
the spatial frequency domain of an observation in a
heterogeneous scheme when using a single-dish for providing 
short-spacings: the inner spatial frequencies are sampled by
the single-dish only, the outer ones are sampled only by the interferometer,
while the overlapping region contains spacings to which both the
single-dish and interferometer are sensitive.}
\end{figure}

We define the calibration scaling factor, $f$, as the ratio of the flux
densities of an unresolved source in the single-dish and interferometer 
maps: 
\begin{equation}
f=\frac{S_{\rm int}}{S_{\rm sd}} \;.
\end{equation}
In the case of perfect calibration, $f=1$. However, $f \neq
1$ otherwise, and needs to be determined very accurately.
Unfortunately, it is hard to find suitable compact sources to 
directly determine $f$.  Hence, the best way to 
estimate $f$ is to compare the
surface brightness of the observed object in the overlap region of the
$u-v$ plane, see Fugure~\ref{f:uv-scheme}. This region should 
correspond to angular sizes to which both
telescopes are sensitive. 
For a source of brightness, $I$, both the interferometer and 
single dish should measure within this
region the same, $I_{\rm sd}=I_{\rm int}$, and calibration errors will
appear as:
\begin{equation}
f=\frac{I_{\rm int}}{I_{\rm sd}} \;.
\end{equation}
For an extended source $I_{\rm int}$ and $I_{\rm sd}$ are often, 
for convenience, expressed in
units of Jy beam$^{-1}$ not Jy sr$^{-1}$, and so will be different numbers
because of the different beams considered (with beam areas $\Omega_{\rm int}$
and $\Omega_{\rm sd}$, respectively).  For this purpose, an estimate of the
resolution difference between the two data sets ($\alpha=\Omega_{\rm
int}/\Omega_{\rm sd}$) is also needed. 

To determine $f$ the following steps are required:
\begin{enumerate}
\item scale the single-dish data by $\alpha$ to account for the difference in
brightness caused {\it only} by different resolutions,
\item Fourier transform the interferometer and scaled single-dish images,
\item deconvolve the single-dish data (by dividing them by the Fourier transform
of the single-dish beam), and  
\item compare `visibilities' in the overlapping region of spatial
frequencies. 
\end{enumerate}

Several important issues should be considered here:
\begin{itemize}
\item When Fourier transforming in Step 2 watch for edge-effects!
To avoid nasty edge-effects in some cases apodizing of both interferometer
and single-dish images may be required in order to make the image
intensities smoothly decrease to zero near the edges.
\item Step 3 requires a very good knowledge of the single-dish beam! To make
things even harder, the FWHM of the single-dish beam and the calibration
scaling factor are highly coupled (Sault \& Killeen 1998). Therefore, 
an error in the 
single-dish beam model has the same effect in the
overlapping region as an error in the flux density scale.
If the single-dish beam is poorly known, $f$ will be a quadratic 
function of distance in the Fourier plane, in the first approximation 
(see Stanimirovic 1999).
\item A sufficient overlap in spatial frequency is required for Step 4.
Assuming a Gaussian-tapered illumination pattern for a single-dish, and
considering a cut-off level of 0.2 for 
reliable data, we 
can estimate the minimum diameter, $D$, of a single-dish necessary 
to provide all spacings shorter than $d_{\rm min}$ for a given interferometer: 
\begin{equation}
D > 1.5 \times d_{\rm min} \;.
\end{equation}
In order to have a reasonable overlap of spatial frequencies so that $f$ can
be derived, a slightly larger single-dish is required with 
$D > 2 \times d_{\rm min}$. For example, for the ATCA shortest 
baseline of 31 m, the single-dish providing short 
spacings should have diameter of $D \geq 62$ m. Therefore, the 64 m Parkes
telescope can do a really great job. 
Also, while the 100-m Green Bank Telescope will be able
to provide short-spacing data for the VLA C and D arrays (with $d_{\rm
min}=35$ m), only Arecibo could do so for the VLA B array (with $d_{\rm
min}=210$ m).     
\end{itemize}

\section{Data Combination in the Spatial Frequency Domain}
\label{s:adding-uv}

\subsection{Theoretically...}

As shown by Bajaja \& van Albada (1979), the true missing short-spacing 
visibilities can be provided from the function $V(u,v)$ in
Equation~\ref{e:sd2}, if the single-dish is large enough to cover the whole
central gap in the interferometer $u-v$ coverage.  The deconvolution of
the single-dish data gives the true single-dish visibilities, where $b_{\rm
sd}(u,v) \neq 0$, by:
\begin{equation}
\label{e:sd3}
V(u,v) = \frac{V_{\rm sd}'(u,v)}{b_{\rm sd}(u,v)} \;.
\end{equation}
Function $V(u,v)$ can be then substituted in Equation~\ref{e:int1}, 
after rescaling by $f$, everywhere in the $u-v$ plane 
where Equation~\ref{e:sd3} holds.
This would provide the resultant $u-v$ coverage having the inner
visibilities from the single-dish data only (rescaled to match
the interferometer flux-density scale due to the calibration differences),
and the outer visibilities from just the interferometer data. 
This is effectively feathering or padding the
interferometer visibilities with the single-dish data.

Depending on the type of input images used, and/or the type of weighting
applied within the region of overlapping spatial frequencies, there are
several applications of this technique (Roger et al. 1984; Vogel et
al. 1984; Wilner \& Welch 1994; Zhou et al. 1996).
However, in all cases, for a good data combination the single-dish 
data set should fulfill the following two conditions.
\begin{itemize} 
\item A sufficiently fine sampling of the single-dish data at the Nyquist
rate (2.4 pixels across the beamwidth) is required to avoid aliasing during
deconvolution (Vogel et al. 1984).  The single-dish data must also have the same
coordinate system as the interferometer data.  Therefore, it is sometimes
necessary to re-grid single-dish images.
\item Visibilities derived from the single-dish data
should have a signal-to-noise ratio comparable to those of
the interferometer in the overlapping region in order not to
degrade the combined map (Vogel et al. 1984).
\end{itemize} 
 
\subsection{Practically...}
\label{s:immerge}

\begin{figure}
\plotfiddle{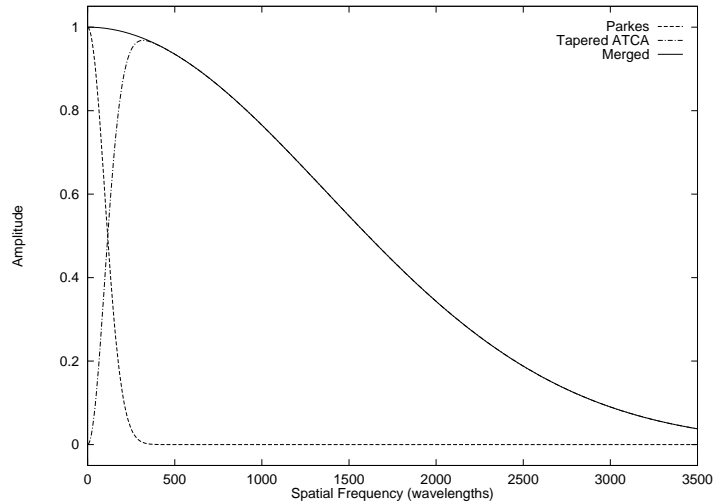}{8cm}{0}{75}{75}{-185}{-5}
\vspace{-1cm}
\caption{\label{f:immerge-beam}The tapering of spatial frequencies sampled
by a single-dish (dashed line), by an interferometer (dot-dashed line) 
and as merged together by the {\sc
miriad} task {\sc immerge} (solid line). The ATCA interferometer 
and the Parkes single-dish are shown as an example (from Sault \& 
Killeen 1998).}
\end{figure}

This linear data combination in the spatial frequency domain is very 
widely used and is implemented in several packages for radio data reduction, 
e.g. task {\sc imerg} in {\sc aips}, {\sc image tool} in {\sc aips++} 
and task {\sc immerge} in {\sc miriad}. As an example we will 
discuss {\sc immerge} here.

{\sc immerge} takes as input a clean (deconvolved) high-resolution image, and a 
non-deconvolved, low-resolution image. These
images are Fourier transformed (labeled as $V_{\rm int}(k)$ and 
$V'_{\rm sd}$) and combined in the Fourier domain applying 
tapering functions, $w'(k)$ and $w''(k)$, such that their sum is 
equal to the Gaussian function having a FWHM of the interferometer, 
$\theta_{\rm int}$:
\begin{eqnarray}
V_{\rm comb}(k) = w'(k)V_{\rm int}(k) + f w''(k)V_{\rm sd}'(k) \\ 
w'(k) + w''(k) = \frac{1}{\sqrt{2\pi}} \exp \left 
(-\frac{\theta_{\rm int}^{2} k^{2}}{4\ln 2} \right ) \;.
\end{eqnarray}
Function $w''(k)$ is a Gaussian with the FWHM of the single dish.  The 
low-resolution visibilities are multiplied by $f$ to account for the
calibration differences.  The final resolution is that of the
interferometer image. The tapering functions $w'$ and $w''$ are shown in
Fugure~\ref{f:immerge-beam}, together with the tapering function of the
merged data set, for the case of the ATCA interferometer and the Parkes
single-dish telescope. {\sc immerge} can also estimate the calibration 
scaling factor, $f$, by comparing single-dish and interferometer data in the
region of overlapping spatial fequencies specified by the user.
 
\begin{figure}
\caption{\label{f:fourier_comb}{\bf [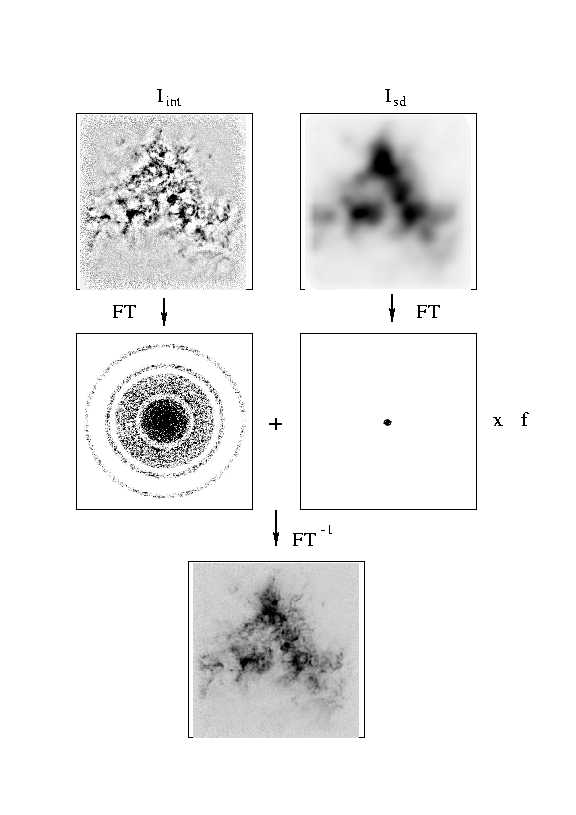]}The 
first row: the ATCA and Parkes images
of the SMC at heliocentric velocity 169 \kms. The second row: amplitudes
of the Fourier transforms of the two input 
images (shown in the first row). Note how the
Fourier transform of the clean interferometer image does not have a hole in
the center, this is the result of extrapolation before
deconvolution. However, the central visibilities are not correctly
represented. The third row:
the output of {\sc immerge} obtained by weighting the distributions from the
second row, combining them, and Fourier transforming 
the result back to the image domain. }
\end{figure}

As an example of how {\sc immerge} works in practice, 
Fugure~\ref{f:fourier_comb} shows a sequence of images at various stages of data
processing: before merging, after Fourier transforming, and the final version.

\section{Data Combination in the Image Domain}
\label{s:comb-image}

There are two distinct methods for data combination in the image
domain. The first, the `linear combination' method (Stanimirovic et
al. 1999), merges data sets in a simple linear fashion before 
final deconvolution, while the second (Sault \& Killeen 1998), 
the non-linear method, combines all data
during the deconvolution process.

\subsection{The `Linear combination' approach}
\label{s:linear}

The theoretical basis for merging before deconvolution is the linear
property of the Fourier transform: a Fourier transform of a sum of two
functions is equal to the sum of the Fourier transforms of the 
individual functions.
Therefore, instead of adding two maps in the Fourier domain and Fourier
transforming the combined map to the image domain, one can produce the same
effect (fill in missing short-spacings in an interferometer $u-v$
coverage) by adding maps in the image domain.  This method was first 
applied by Ye \& Turtle (1991), and Stewart et al. (1993).

As we have seen from Equations~\ref{e:int} and \ref{e:sd1}, the 
interferometer and single-dish data obey the 
convolution relationship.
The dirty images and beams can be combined to form a composite
dirty image ($I^{D}_{\rm comb}$) and a composite beam ($B_{\rm comb}$)
with the following weighting:
\begin{eqnarray}
\label{e:comb}
I^{D}_{\rm comb}=(I^{D}_{\rm int} + \alpha f I_{\rm
sd}^{\rm D})/(1+\alpha) \\
B_{\rm comb}=(B_{\rm int} + \alpha B_{\rm sd})/(1 +\alpha),
\end{eqnarray}
where $\alpha=\Omega_{\rm int}/\Omega_{\rm sd}$ is an estimate of the
resolution difference between the two data sets.
The convolution relationship, $I^{D}_{\rm comb} = B_{\rm
comb} * I$, still exists between the composite dirty image, $I^{D}_{\rm
comb}$, and the true sky brightness distribution, $I$.
Deconvolving the composite dirty image with the composite beam hence solves for
$I$.

\begin{figure}
\vspace{0.5cm}
\caption{\label{f:image_comb}{\bf [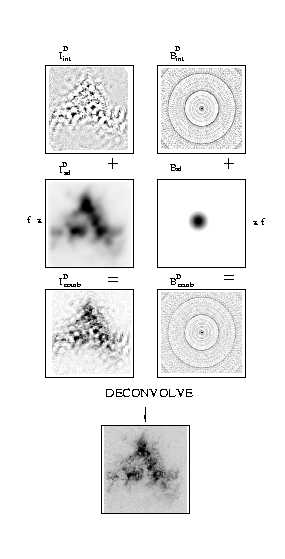]}The first 
row (top): the ATCA `dirty' image 
of the SMC at 169 \kms, and the ATCA `dirty' beam for a single pointing. 
The second row: the Parkes image
of the SMC at the same heliocentric velocity, and the Parkes beam assumed to be
a Gaussian function with the FWHM of 18.8 arcmin. The third row: 
$I^{D}_{\rm comb}=(I^{D}_{\rm int} + \alpha f I_{\rm sd}^{\rm
D})/(1+\alpha)$ and $B_{\rm comb}=(B_{\rm int} + \alpha B_{\rm sd})/(1
+\alpha)$, where $\alpha = 7.5\times10^{-3}$. 
The fourth row (bottom): the result of the MEM 
deconvolution of $I^{D}_{\rm comb}$ by $B_{\rm comb}$.}
\end{figure}

There is no single existing program (task) that employs 
this method, but a linear
combination of maps can be easily obtained in any package for radio data
reduction, followed by a favorite choice of a deconvolution algorithm.
As an example, Fugure~\ref{f:image_comb} shows the `linear combination' method
applied on the ATCA mosaic and Parkes telescope HI observations of the 
SMC at several stages of data processing. 
Note that $B_{\rm int}$ in the case of mosaic observations represents
a whole cube of beams, one for each pointing in the mosaic. The combined dirty
image was deconvolved using {\sc miriad}'s maximum entropy algorithm
(Sault, Staveley-Smith, \& Brouw 1996).  The model was finally restored
with a 98-arcsec Gaussian function.

\subsection{Merging during deconvolution}

\begin{figure}
\caption{\label{f:mem}{\bf [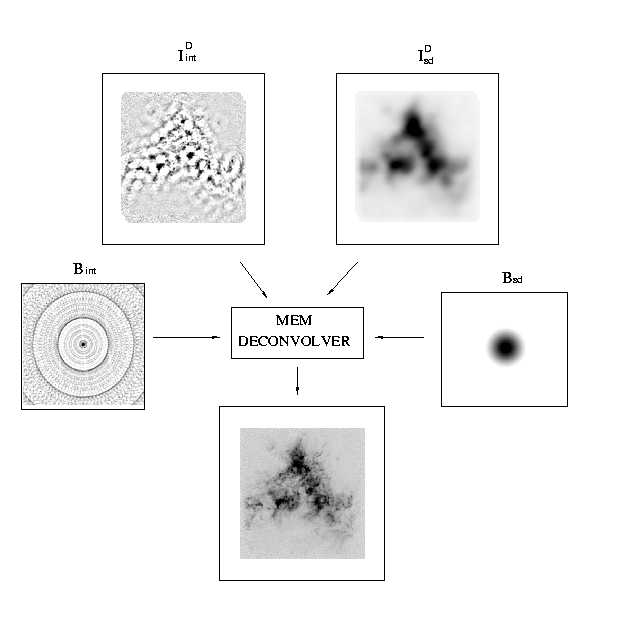]}A schematic example 
of the non-linear approach for
adding short spacings performed by {\sc miriad}'s program {\sc mosmem}. 
In the {\bf `default' image} method the low-resolution 
Parkes image of the SMC is used to constrain missing spatial
frequencies in the center of the ATCA-image $u-v$ plane. In 
the {\bf `joint' deconvolution} method, the model is found by fitting both
data sets simultaneously while maximizing the entropy.}
\end{figure}

Besides the missing information in the center of the $u-v$ plane, an
interferometer $u-v$ coverage suffers from spatial-frequency gaps. Since
the missing information can be introduced in an infinite number of ways,
the convolution equation ($I^{D}=I*B$) has a non-unique solution.  Hence,
the deconvolution has the task of selecting the `best' image from all those
possible (Cornwell 1988). Since deconvolution has to estimate
missing information, a non-linear algorithm must be employed
(Cornwell 1988; Sault et al. 1996).  Cornwell (1988) and
Sault et al. (1996) showed that the deconvolution 
algorithms implementing the
so called `joint' deconvolution scheme, whereby observations of many
pointing centers are combined prior to deconvolution,
 produce superior results in the case of
mosaicing, since more information is fed to the deconvolver. We expect
that the same argument might apply for the addition of the single-dish 
data, resulting in the merging before and during deconvolution being 
more successful than the merging of clean images in the spatial 
frequency domain.
 
The maximum entropy method (MEM) is one of the non-linear deconvolution
algorithms.  It selects the deconvolution solution so it fits the data and,
at the same time, has a maximum `entropy'. Cornwell (1988) explains this
entropy as something which when maximised produces a positive image with a
compressed range in pixel values. The compression criterion forces the
final solution (image) to be smooth, while the positivity criterion forces
interpolation of unmeasured Fourier components.  One of the commonly used
definitions of entropy is:
\begin{equation}
\label{e:mem}
\aleph = - \sum_{i} I_{i} \ln \left (\frac{I_{i}}{M_{i}e} \right )
\end{equation}
where $I_{i}$ is the brightness of $i$'th pixel of the MEM image and
$M_{i}$ is the brightness of $i$'th pixel of a `default' image incorporated
to allow {\it a priori} knowledge to be used ($e$ is the base of the natural
logarithms).  The requirement that the final image fits the data is usually
incorporated in a constraint such that the fit $\chi^{2}_{\rm int}$ of the
predicted visibilities to those observed (Cornwell 1988) is close to
the expected value:
\begin{equation}
\chi^{2}_{\rm int} \leq N \sigma_{\rm int}^{2},
\end{equation}
with $N$ being the number of independent pixels in the map and $\sigma_{\rm
int}^{2}$ being the noise variance of the interferometer data.
 
The single-dish data can be incorporated during the maximum-entropy
deconvolution process in two ways.  
 
\begin{enumerate}
\item {\bf The `default' image }\\
The easiest way is to use the single-dish data as a `default' image in
Equation~\ref{e:mem} since, in the absence of any other information or
constraints, this forces the deconvolved image to resemble the single-dish
image in the spatial frequency domain where the interferometer data
contribute no information.  Since this method puts more weight on the
interferometer data wherever it exists, the size of the overlapping region
plays a very important role (Holdaway 1999).  As large an overlap of
spatial frequencies as possible is required to provide good quality
interferometer and single-dish data within this region, in order to retain
the same sensitivity over the image. 
 
\item {\bf The `joint' deconvolution}\\
The second way maximizes the entropy while being
subject to the constraints of fitting both data sets simultaneously:
\begin{eqnarray}
\aleph = - \sum_{i} I_{i} \ln \left (\frac{I_{i}}{e} \right ) \\
\sum_{i} \left \{I^{D}_{\rm int} - B_{\rm int}*I \right \}_{i}^{2} <  
N \sigma_{\rm int}^{2} \\
\label{e:statistics}
\sum_{i} \left \{I^{D}_{\rm sd} - \frac{B_{\rm sd}*I}{f} 
\right \}_{i}^{2} <  M \sigma_{\rm sd}^{2} \;. 
\end{eqnarray}
The `joint' deconvolution method provides also an alternative
way, completely performed in the image domain, for determining the
calibration scaling factor. Maximizing the entropy, while fitting both data
sets, a `joint' deconvolution algorithm can iteratively
solve for $I$ and $f$ simultaneously. 
\end{enumerate}

As a schematic example, Fugure~\ref{f:mem} shows the non-linear approach for
data combination performed by {\sc miriad}'s program {\sc mosmem}.

\section{Comparison of the Different Methods}
\label{s:comparison}

\begin{figure}
\caption{\label{f:4methods} {\bf [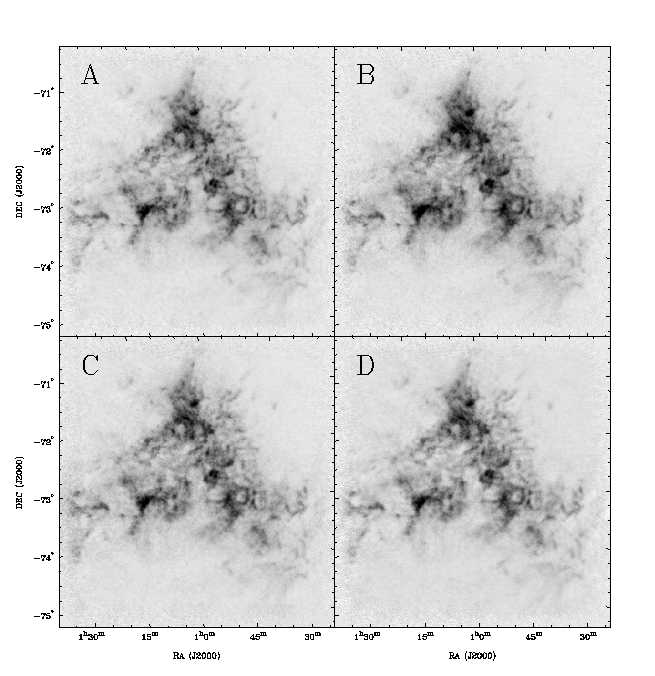]}The 
result of adding Parkes short-spacings 
into an ATCA
HI mosaic of the SMC at heliocentric velocity 169 \kms.  Four different
methods are shown from top-left to bottom-right: data combination in the
Fourier domain using {\sc immerge} (panel A), the `linear
combination' (panel B), data combination during deconvolution using
Parkes data as a `default' image (panel C) and the `joint' deconvolution
of both Parkes and ATCA data (panel D). All images have grey-scale
range $-11$ to 107 K with a linear transfer function. }
\end{figure}

The qualitative comparison of all four methods for the short-spacings
correction addressed in Sections~5 
and~6, is shown in
Fugure~\ref{f:4methods} for the 169 \kms~velocity channel of the SMC
data. All four images have the same grey-scale range ($-11$ to 107 K) and
are remarkably similar.  They all have the same resolution and show the
same small and large scale features with only slightly different flux-density
scales. No signs of interferometric artifacts are visible on any of the
images. This demonstrates that all four methods for the 
short-spacings correction
give satisfactory results in the first approximation, when an {\it a
priori} determined calibration scaling factor is used. 
A similar conclusion was reached in Wong \& Blitz (2000) where
results of data combination using {\sc immerge} and the `linear
combination' method (Section~\ref{s:linear}) were compared for the 
case of BIMA and Kitt Peak 12-m telescope CO observations.

{\small
\begin{table}
\begin{center}
\caption
{\label{t:comparison} Total flux density, minimum and maximum brightness
values and noise for the SMC
ATCA image at 169 \kms~corrected for Parkes short-spacings. Results of four
different methods are shown: data combination in the Fourier domain using
{\sc immerge} (labeled as method A), `linear combination' (labeled as
method B), data combination during deconvolution using Parkes data as a
`default' image (labeled as C) and the `joint' deconvolution of both Parkes
and ATCA data (labeled as D).  The total flux density 
of the Parkes image alone is 6100 Jy.}
\vskip .1truein
\begin{tabular}{ccccc}
\tableline
Method & Total Flux & Min & Max & Noise \\ & (Jy) & (Jy beam$^{-1}$) & (Jy
       beam$^{-1}$) & (mJy beam$^{-1}$) \\

\tableline
A& 5600 & $-0.24$ & 1.97 &  30  \\
B& 6500 & $-0.21$ & 2.16 &   32  \\
C& 6300 & $-0.21$ & 2.00 &    28  \\
D& 5900 & $-0.30$ & 2.00& 29  \\
\tableline 
\tableline
\end{tabular}
\end{center}
\end{table} 
}

The quantification of the quality of an image depends on the
scientific questions we want to address (Cornwell et al. 1993) and is
therefore case specific. 
Something that any short-spacings correction must fulfill, though, is that
the resolution of the final image should be the same as for the interferometer
data alone, while the integrated flux density 
of the final image should be the same
as measured from the single-dish data alone.  Table~\ref{t:comparison} shows
measurements of the total flux density, minimum/maximum values and noise
level in the four resultant images. 
All four images have very comparable noise levels, minimum values and
maximum values. The last two images also have very comparable (within 3\%)
total flux density, relative to the Parkes value alone, while the first two have
lower (by 8\%) and higher (by 7\%) flux densities, respectively.  The
differences come, most likely, from the different weighting of the single-dish
data employed by the different methods. While {\sc immerge} slightly
over-weights very short interferometric spatial frequencies, the `linear
combination' method slightly over-weights single-dish data in the region of
overlap. This results in the total flux density being slightly lower in the
first case, and slightly higher in the second case.

A few (general) remarks on all four methods: 
\begin{itemize}
\item The `feathering' or Fourier method (A) is the fastest and the 
least computer intensive way to add short-spacings.  It is also the most
robust way relative to the other three methods which all require a
non-linear deconvolution at the end. 

\item The great advantage of the `linear combination' method is that it does
not require either Fourier transformation of the single-dish data, 
which can suffer severely from edge effects, nor deconvolution of 
the single-dish data which is especially uncertain and  leads to 
amplification of errors.

\item The `default' image method shows surprisingly reliable results 
when a significantly large single-dish is used.

\item Adding during deconvolution when fitting both data sets 
simultaneously provides, theoretically, the best way 
to do the short-spacing correction. 
However, this method depends heavily on a good estimate
of the interferometer and single-dish noise variances.

\end{itemize}

\section{Summary}

In this article the need  for, and methods of combining interferometer and
single-dish data have been explained and demonstrated.
This combination is an important step when mapping extended objects and it is 
becoming a standard observing and data processing technique.

After a brief introduction to interferometry in
Section~2 the short-spacings problem 
and general approaches for its solution were
discussed in Section~3. 
To {\it fully and accurately} fill in the missing short-spacings to
interferometric data, a heterogeneous array scheme seems to be
preferable. A sufficiently large single-dish, with diameter at least 1.5 
greater larger than the shortest interferometer baseline, provides the simplest
option. In order to cross-calibrate single-dish and interferometer
data sets, a significant overlap of spatial frequencies is
required. Four different combination methods, 
two linear and two non-linear, for the
short-spacings correction have been discussed and the 
results of applying these methods to
the case of HI observations of the SMC presented. Linear methods
are data combination in the spatial frequency domain and the `linear
combination' method, while data combination during deconvolution provides
two non-linear methods. All four techniques yield satisfactory and comparable 
results. 

\acknowledgments
Many thanks to Darrel Emerson, Chris Salter and John Dickey 
for reading the article and providing valuable suggestions for improvement.
I am also grateful to Matthew Wyndham for his help with figures, 
as well as assisting with last minute crises in organizing the meeting.

\end{document}